# Segmentation of the Left Ventricle by SDD double threshold selection and CHT

*WenFei Zhang Xiufeng Li and ZhenZhou Wang*, Member, IEEE*
College of Electrical and Electronic Engineering,
Shandong University of Technology, China, 255049.
*Corresponding e-mail: wangzz@sdut.edu.cn

## ABSTRACT

Automatic and robust segmentation of the left ventricle (LV) in magnetic resonance images (MRI) has remained challenging for many decades. With the great success of deep learning in object detection and classification, the research focus of LV segmentation has changed to convolutional neural network (CNN) in recent years. However, LV segmentation is a pixel-level classification problem and its categories are intractable compared to object detection and classification. In this paper, we proposed a robust LV segmentation method based on slope difference distribution (SDD) double threshold selection and circular Hough transform (CHT). The proposed method achieved 96.51% DICE score on the test set of automated cardiac diagnosis challenge (ACDC) which is higher than the best accuracy reported in recently published literatures.

## 1. INTRODUCTION

SEGMENTATION of the left ventricle in MRI is important for clinical diagnosis, heart disease monitoring and heart treatment planning [1]. For instance, the volume or the mechanical desynchrony of the left ventricle could be computed for cardiac function analysis. To compute the volume or the mechanical desynchrony accurately and timely, robust and automatic LV segmentation method is indispensible. The challenges of fully automated LV segmentation include: (1), the intensities of the LV and its adjacent myocardial walls vary greatly in the different images, which makes traditional threshold selection methods [2-10] fail in finding the optimum threshold for all the images. (2), the intensities of the LV itself in the same image also vary greatly, which makes many well-known image segmentation methods fail in getting an accurate segmentation result [11-17]. (3), not all the imaged LVs are circular because of the papillary muscles in the myocardium walls, which makes post-processing methods indispensible. (4), the motion of the LV in different frames is not well predictable, which adds difficulty to the tracking based methods [18-19].

In recent years, CNN has almost monopolized the medical image analysis fields and has become the most widely used method for cardiac segmentation [20-27]. Many researchers have claimed that their CNN based methods have achieved state of the art accuracy in their papers. For instance, 94% DICE score has been achieved in 2016 on the medical image computing and computer assisted intervention (MICCAI) 2009 dataset [25], where the deep learning was used to yield an approximation prediction and then deformable model was used to refine the final segmentation result. The method proposed in [25] indicates that CNN methods do not have the ability to achieve an accurate image segmentation result. Thus, the authors choose to optimize the predicted results by deep learning with the evolvement of the deformable model. Later, the reported DICE score by CNN on the MICCAI 2009 dataset dropped to 91% in 2018 [26] and the reported DICE score by CNN on the MICCAI 2009 dataset dropped to 0.9 in 2019 [27]. It is worth noting that our proposed SDD based LV segmentation method achieved 92.46% DICE score on the MICCAI 2009 dataset in 2019 [28] and achieved 94.97% DICE score on the MICCAI 2009 dataset in 2020 [29]. The major reason that makes the CNN based methods the most popular method is that CNN based methods frequently win the international competitions. For instance, CNN based method has also become the winner of the LV segmentation challenge with the ACDC dataset [30]. However,many image segmentation experts have indicated that CNN could only yield a prediction map instead of an accurate segmentation result.

In this paper, we propose an automatic and robust LV segmentation method based on our previous work [29]. The proposed method combines the strength of two techniques. The first one is SDD threshold selection [31] and the second one is circular Hough transform (CHT) [32-33]. SDD threshold selection is used to obtain the accurate pixel-level image segmentation at first. Then CHT is used to separate the LV from the adhesive parts, such as the RV at the base part of the heart. The proposed method is efficient, robust and does not need any time-consuming and tedious training work. Different from our previous studies [28-29], double thresholds are selected by SDD because two different classes of pixels are contained in the LV. The pixels of the first class are bright and generated by blood inside the LV. The pixels of the second class are dark and generated by muscles inside the LV.

## 2. THE PROPOSED METHOD

The flowchart of the proposed method is shown in Fig. 1. and it contains the following eight steps.

**Step 1:** the ROIs are generated for each slice of the patient case based on coarse SDD threshold selection and region growing as described in [29].

**Step 2:** The valley positions at both sides of the first true peak are selected as the double thresholds that are used to segment the LV.

**Step 3:** An inverse operation is used to get the segmented myocardium from the LV segmentation image.

**Step 4:** CHT is used to find the biggest circle in the myocardium segmentation image.

**Step 5:** the biggest circle is binarized and dilated to be an mask image.

**Step 6:** the mask image is used to get the LV from the LV segmentation image by an intersection operation.

**Step 7:** the convex hull of the LV is generated for accuracy improvement.

**Step 8:** the boundary of the LV is extracted from the generated convex hull.

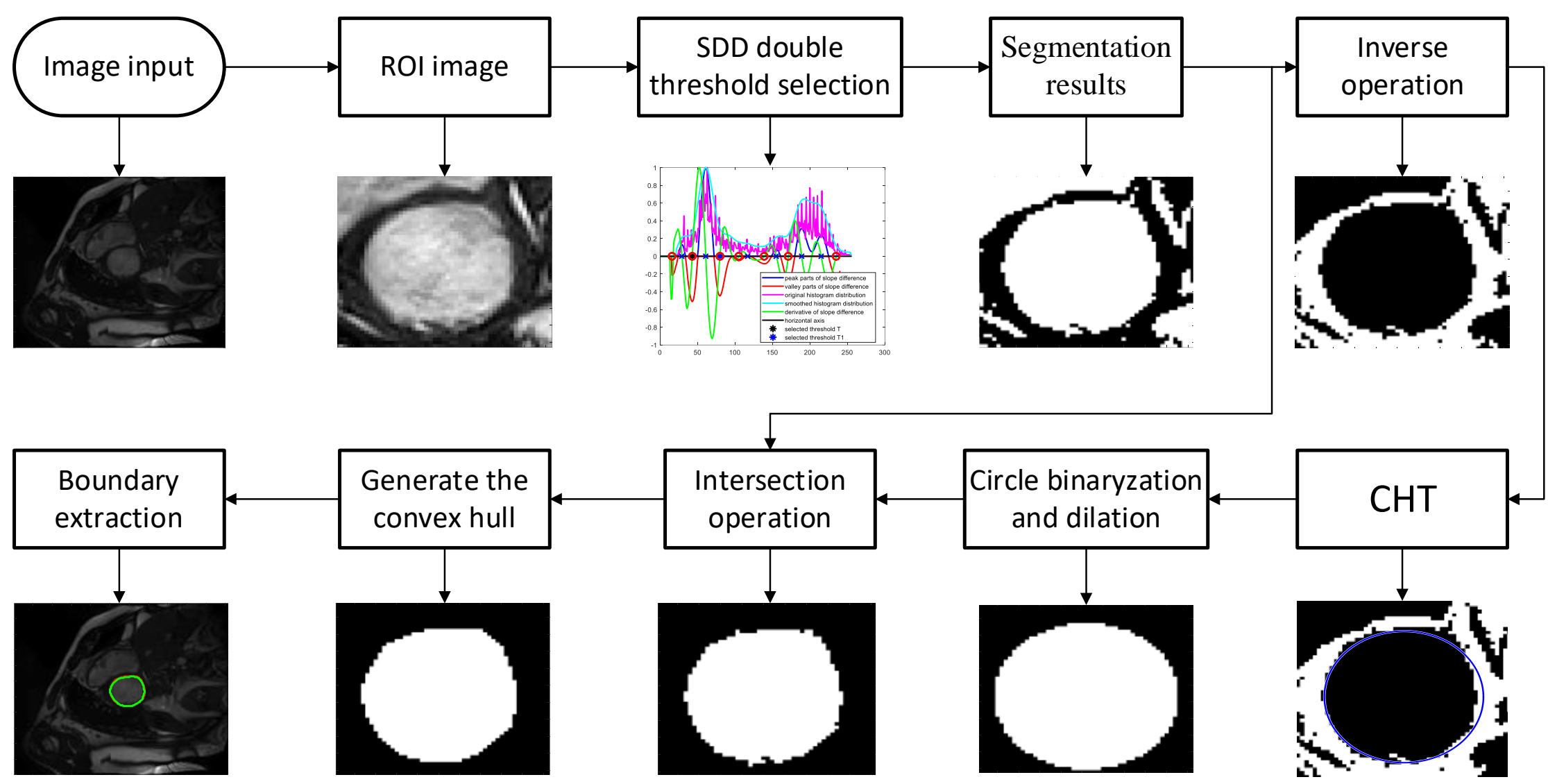

**Fig.1. Flowchart of the proposed LV segmentation method**

## *2.1 LV segmentation by SDD double threshold selection*

From the MR images especially the images at the end systole (ES) stage, we can see that most parts of the LV are bright and some parts of it are dark. The bright parts are generated by blood inside the LV and the dark parts are generated by muscles inside the LV. For robust segmentation of the LV, both the bright part and the dark part should be segmented. In the histogram of the selected ROI, the pixels of the myocardium are distributed in a narrow span on the left while the pixels of the left ventricle are distributed in a wide span on the right. Thus, it is very easy for SDD to distinguish them. The largest true SDD peak must correspond to the clustering center of the myocardium pixels according to the property of SDD [31]. There are two valleys that are on the two sides of the largest SDD peak. The position of the left valley corresponds to the low threshold $T_1$ and the position of the right valley corresponds to the high threshold $T_2$. The pixels of the dark parts are distributed on the left of $T_1$. The pixels of the myocardium are distributed in the interval of $T_1$ and $T_2$. The pixels of bright parts are distributed on the right of $T_2$. Thus, the LV is segmented as follows.

$$L_V = R \le T_1 \,||\, R \ge T_2 \qquad (1)$$

where $L_V$ denotes the LV segmentation result and $R$ denotes the selected ROI. Fig.2 demonstrates the process of LV segmentation by SDD double threshold selection without adhered RV or myocardium. As can be seen from the segmented LV in Fig. 2 (c), some dark parts that belong to the muscles inside the LV are not segmented robustly because their pixels are similar to the those of the myocardium. However, these black holes inside the LV will not affect the accuracy of the LV boundary extraction in the following steps.

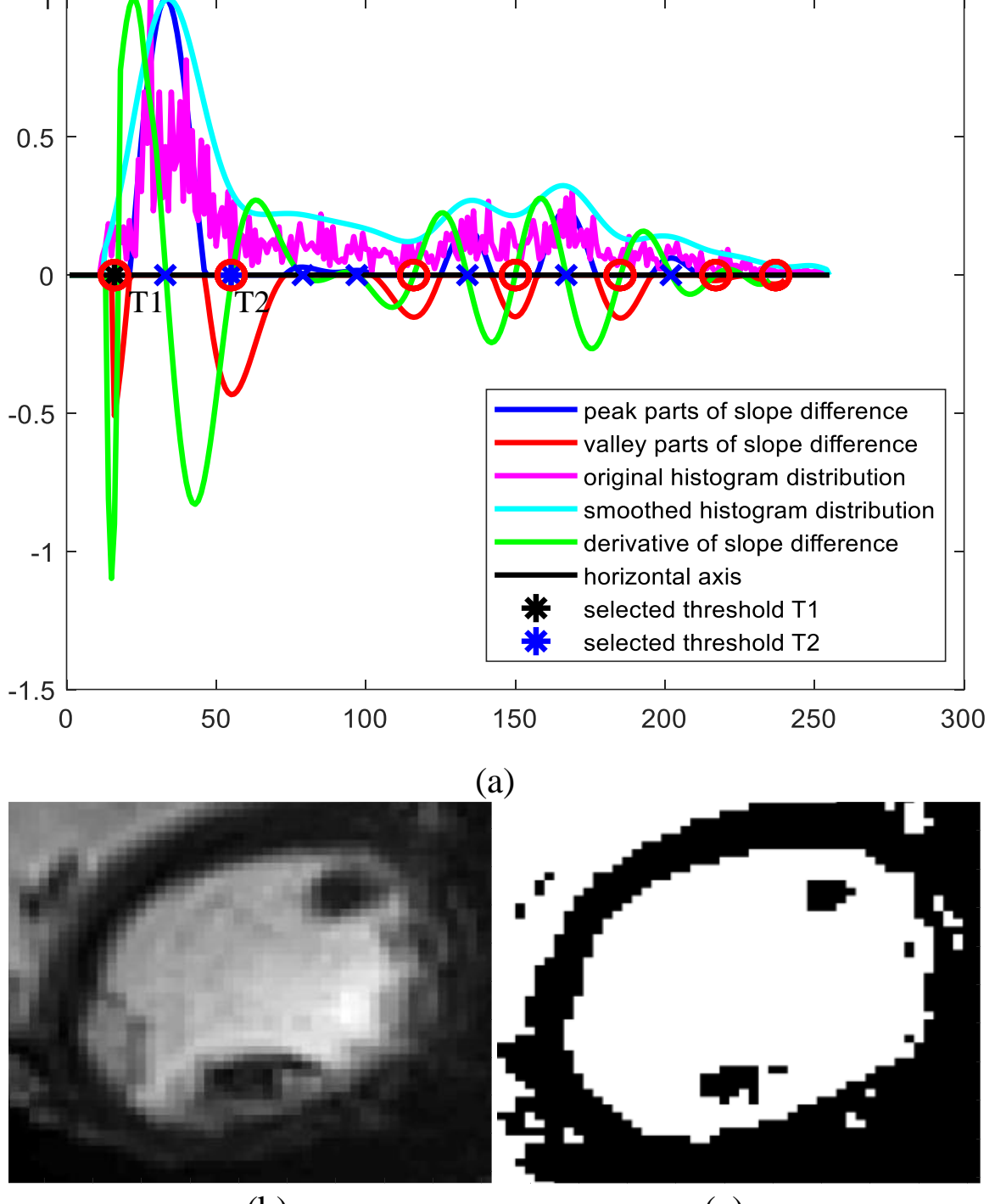

**Fig.2. Demonstration of LV segmentation by SDD double threshold selection.** (**a**) The SDD double threshold selection process; (**b**) The ROI image; (**c**) The LV segmentation result.

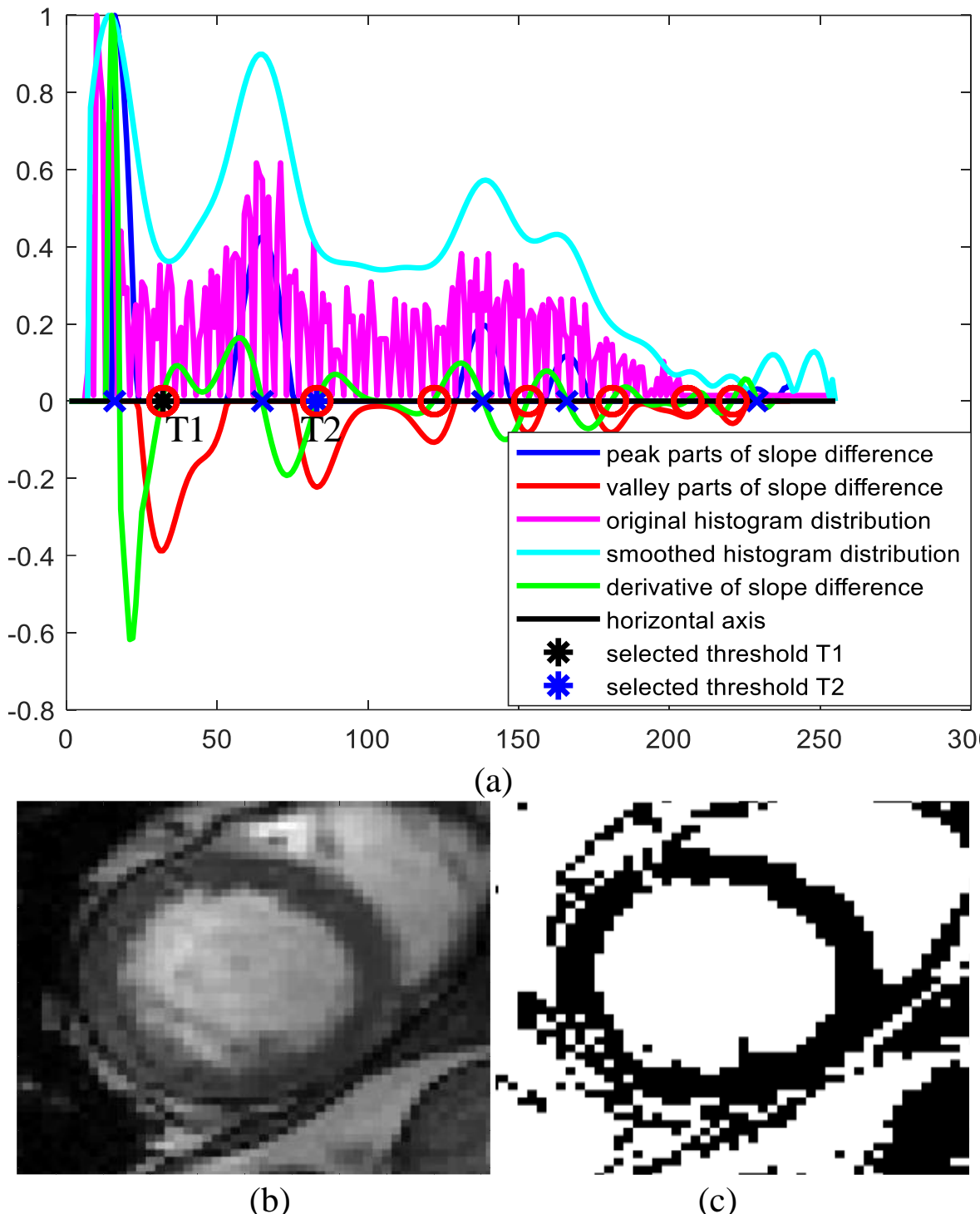

**Fig.3. Demonstration of LV segmentation by SDD double threshold selection.** (**a**) The SDD double threshold selection process; (**b**) The ROI image; (**c**) The LV segmentation result.

Fig. 3 demonstrates the process of LV segmentation by SDD double threshold selection with adhered RV or myocardium. As can be seen from the segmented LV in Fig. 2 (c), the LV is adhered to the myocardium. If the LV segmentation accuracy is computed directly, the segmentation error will be very large. If a morphological erosion is used to separate the LV from the myocardium, the segmentation accuracy will still be affected. To resolve this issue, the following improvement method based on the CHT detected circle is proposed.

### *2.2 Improvement of the LV segmentation result based on the CHT detected circle*

After the LV is segmented by double SDD thresholds, it may be adhered to the right ventricle or the myocardium. Thus, CHT is used to find a circle to separate the LV from the adhered RV or myocardium. This is hinted by the fact that the LV is similar to the circle. Although it is most similar to a ellipse than a circle, the process of finding the ellipse is much more time consuming. Thus, CHT is used instead of ellipse Hough transform. Because CHT is more accurate in finding circles from the inner boundaries instead of the outer boundaries, we transform the outer boundary of the LV into inner boundary by an inverse operation which is formulated by the following equation:

$$M_S = \sim L_V \tag{2}$$

where $M_S$ denotes the myocardium segmentation result.

Then CHT is used find the circles inside the inner boundaries. When more than one circle is found, the biggest one is selected. The circle is binarized and dilated once by a disk structure element with radius 3.

$$C_D = C_B \oplus B \tag{3}$$

where $C_B$ denotes the binarized circle and $C_D$ denotes the dilated circle. $B = \{(0,1,0;1,1,1;0,1,0)\}$ is the disk structure element. An intersection operation is then conducted between the LV segmentation result $S_{LV}$ and the dilated circle $C_D$.

$$L_I = L_V \cap C_D \tag{4}$$

where $L_I$ denotes the binary LV selected by the dilated circle. The selected binary LV is morphologically filtered to remove small blobs and then post-processed by a convex hull operation to compensate for the errors caused by papillary muscles.

In Fig. 4, the same image used to demonstrate the effectiveness of the SDD double threshold selection in Fig. 3 is used to demonstrate the effectiveness of the improvement effect of the CHT detected circle. Fig. 4 (a) shows the detected CHT circle superimposed on the myocardium segmentation result. Fig. 4 (b) shows the dilated binary circle. Fig. 4 (c) shows the LV segmentation result obtained by Eq. (1). Fig. 4 (d) shows the intersection of the LV segmentation result and the dilated binary circle. As can be seen, the LV is separated from the adhered myocardium successfully.

In Fig. 5, a typical image from the base part of a slice is used to demonstrate the effectiveness of the proposed segmentation accuracy improvement method based on CHT detected circle further. Fig. 5 (a) shows the detected CHT circle superimposed on the myocardium segmentation result. Fig. 5 (b) shows the dilated binary circle. Fig. 5 (c) shows the LV segmentation result obtained by Eq. (1). Fig. 5 (d) shows the intersection of the LV segmentation result and the dilated binary circle. As can be seen from Fig. 5 (c), the segmented LV is adhered to the RV which will decrease the segmentation accuracy significantly. From Fig. 5 (d), we can see that the LV is separated from the adhered RV successfully based on the CHT detected circle.

In Fig. 6, we demonstrate why we use the myocardium segmentation result instead of the LV segmentation result to detect the CHT circle. As can be seen, multiple CHT circles will be found when the LV segmentation result is used, which will complicate the selection process, which will definitely increase the processing time and might decrease the final segmentation accuracy slightly.

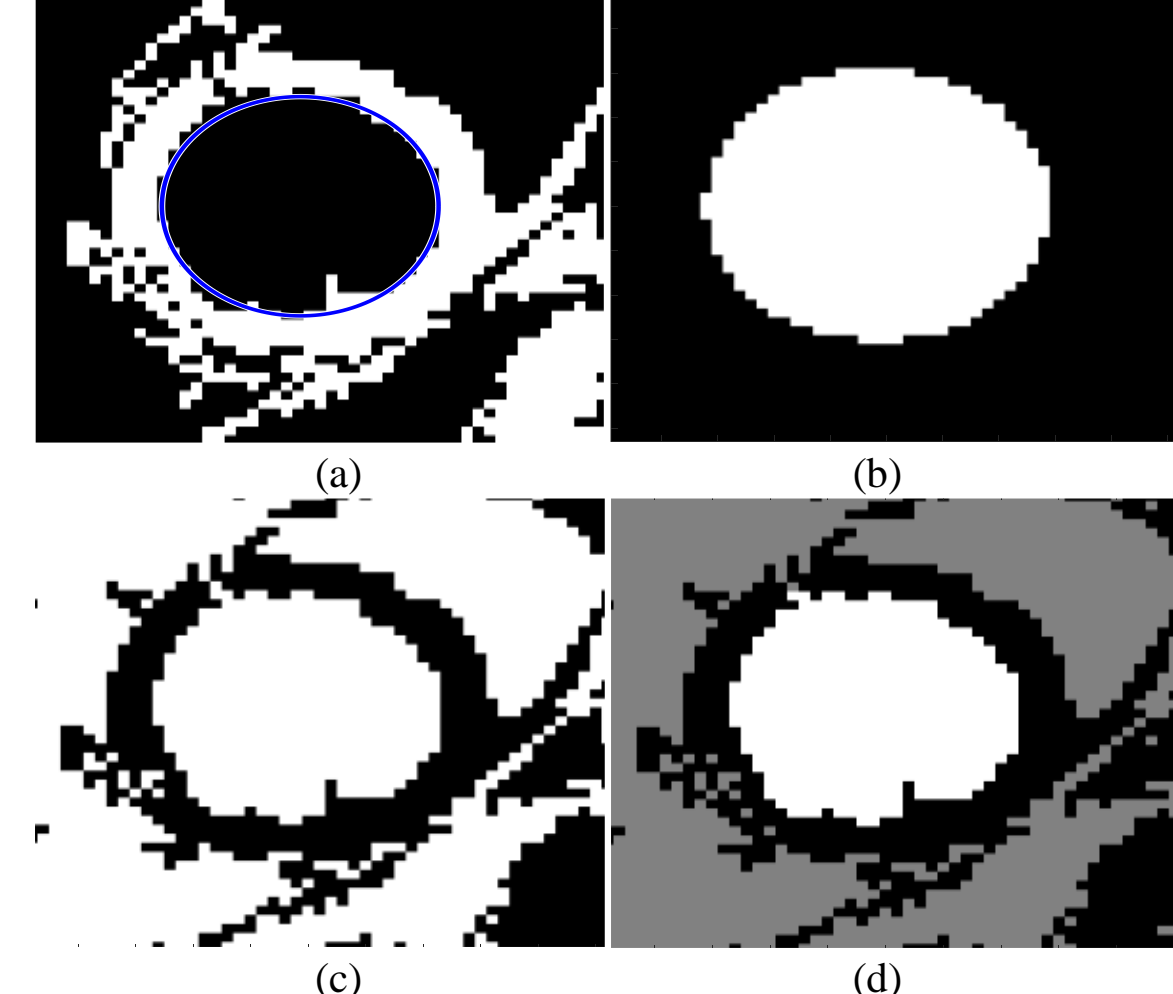

**Fig.4. Demonstration of the improvement process of the LV segmentation result by CHT.** (**a**) The CHT circle superimposed on the myocardium segmentation result; (**b**) The dilated binary circle; (**c**) The LV segmentation

result; (**d**) The intersection part of the LV segmentation result with dilated binary circle (denoted in white).

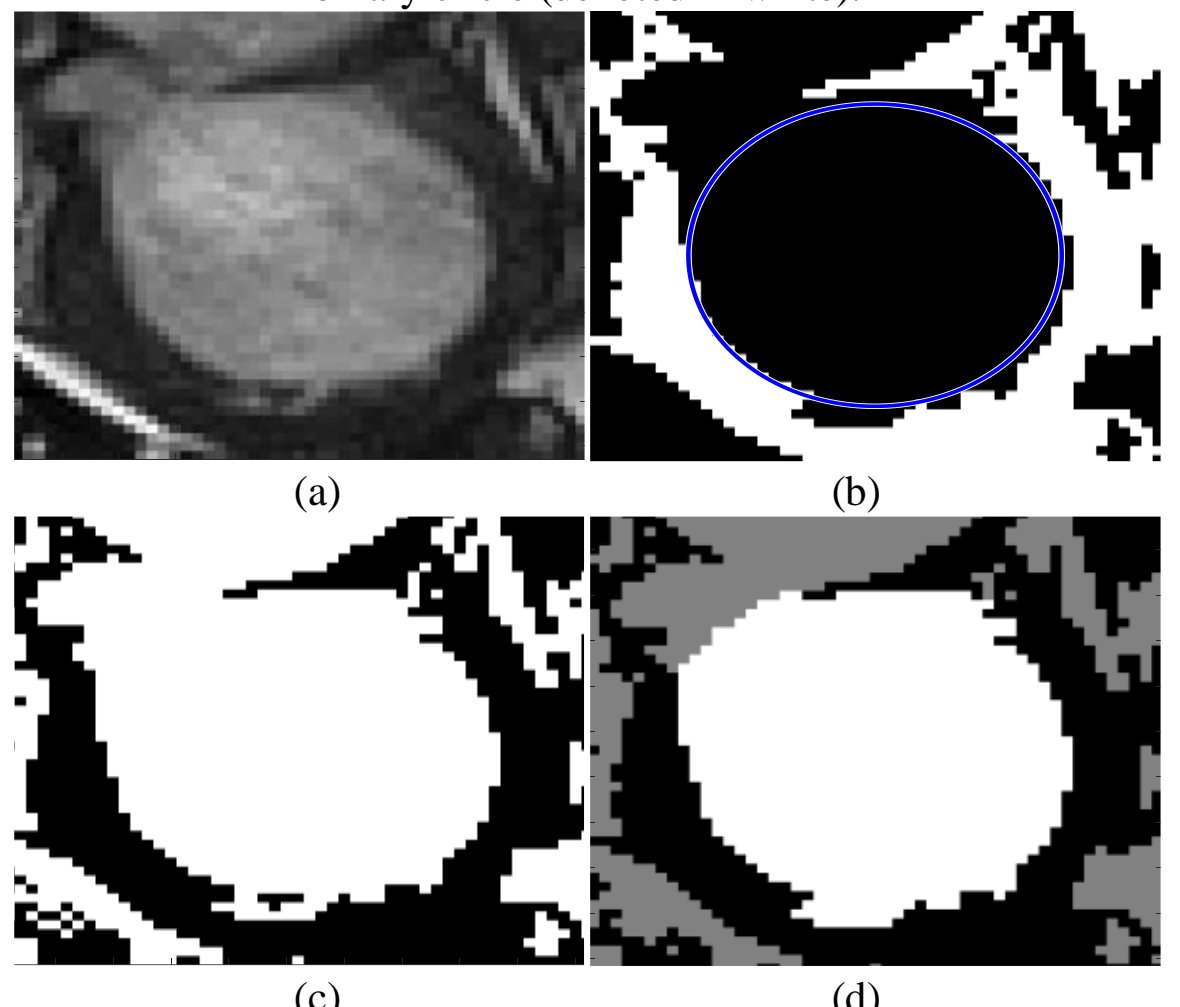

(a) (b)

(c) (d)

**Fig.5. Demonstration of the improvement method with a typical image from the base part of the heart.** (**a**) The ROI image; (**b**) The CHT circle overlaying on myocardium segmentation result; (**c**) The LV segmentation result; (**d**) The intersection of the LV segmentation result with dilated binary circle (denoted in white).

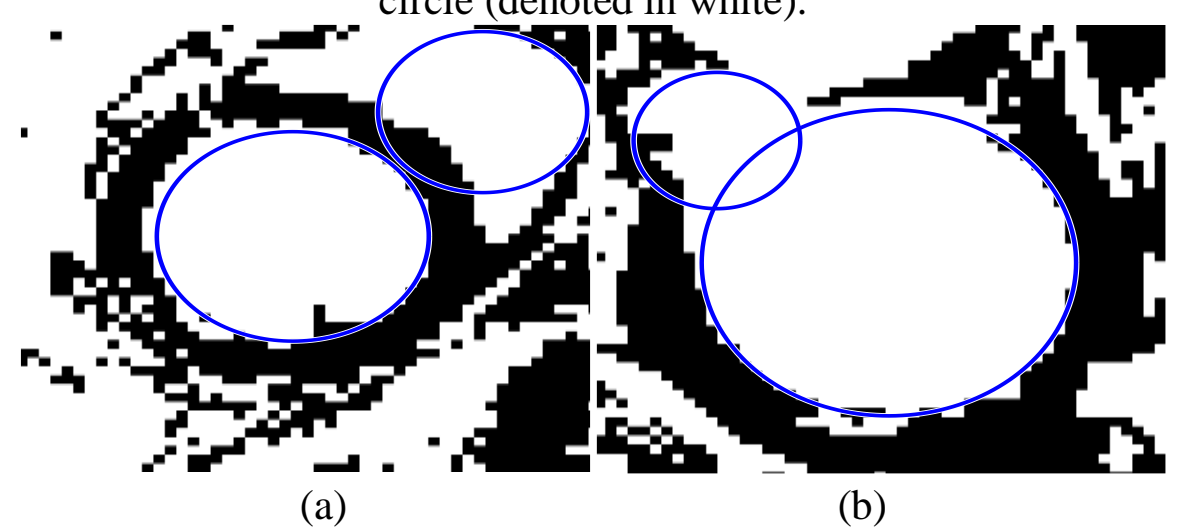

(a) (b)

**Fig.6. Demonstration of detecting the CHT circles based the LV segmentation result.** (**a**) The detected CHT circles for the first demonstration image; (**b**) The detected CHT circles for the second demonstration image.

## 3 EXPERIMENTAL RESULTS

The ACDC dataset is used to test the proposed method compare it with state of the art methods [34-39]. There are 90 patient cases in total in ACDC dataset for both training and testing. The test dataset includes patient007, patient009, patient018, patient035, patient041, patient042, patient052, patient067, patient071, patient075, patient084 and patient088. Since the proposed method does not require any training datasets, we did not use the patient cases in the training set in this study.

Two accuracy measures are used for quantitative comparison. The first measure is DICE score and the second measure is Hausdorff distance. For the DICE score, the higher the better. For the Hausdorff score, the smaller the better. The quantitative accuracy comparisons between the proposed method and state of the art methods are shown in Table 1. As can be seen, the proposed method achieved the highest DICE score and achieved the smallest Hausdorff score.

We show some qualitative results of the proposed method in Fig. 7 and Fig. 8 for visual comparisons with those of state of the art methods. As can be seen, the proposed method is accurate enough for the clinical usage. Actually, the errors between the manually generated ground truth boundary and the automatically generated boundary are caused by the noise instead of the method's defect.

Table 1. Quantitative comparison of the segmentation accuracy of the proposed method with those of state of the art CNN methods (**the bold value denotes the best**).

| Methods | DICE | Hausdorff |
| --- | --- | --- |
| [29] | 0.9497 | 2.24 |
| [34] | 0.915 | 2.524 |
| [35] | 0.958 | 3.43 |
| [36] | 0.918 | NA |
| [37] | 0.92 | 10.45 |
| [38] | 0.92 | 6.2 |
| [39] | 0.9375 | 9.247 |
| **Proposed** | **0.9651** | **2.12** |

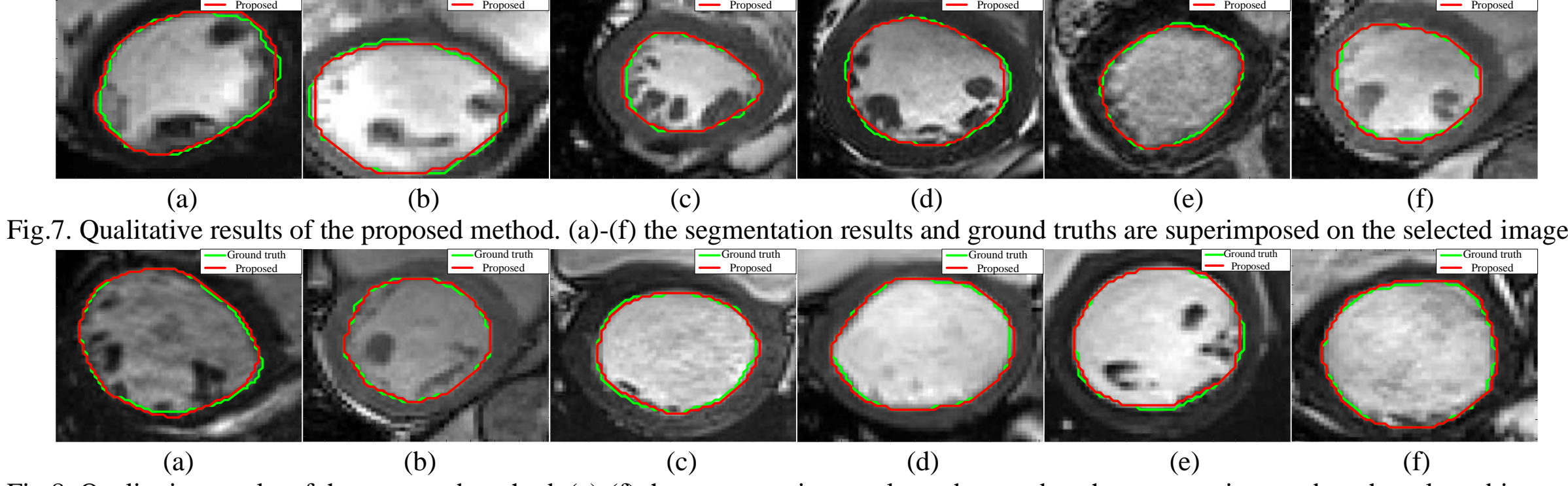

(a) (b) (c) (d) (e) (f)

Fig.7. Qualitative results of the proposed method. (a)-(f) the segmentation results and ground truths are superimposed on the selected images.

(a) (b) (c) (d) (e) (f)

Fig.8. Qualitative results of the proposed method. (a)-(f) the segmentation results and ground truths are superimposed on the selected images.

## 4 CONCLUSION

In this paper, a double-threshold based LV segmentation method is proposed to extract the boundary of the LV from the magnetic resonance images robustly. The double thresholds are calculated by SDD robustly and then used to segment the LV automatically. CHT is used to detect a circle to remove the RV part or myocardium part adhered to the LV. Experimental results verified both the SDD double threshold selection method and the CHT based accuracy improvement method. We compared the proposed method with state of the art methods on the publicly accessible ACDC dataset. The proposed method achieved the highest DICE score and the smallest Hausdorff score among all methods published in recent literatures.